\documentclass[journal=jpcl,manuscript=article]{achemso}
\usepackage{graphicx}
\usepackage{dcolumn}
\usepackage{bm}
\usepackage{rotating} 
\usepackage{epsfig}
\usepackage[version=3]{mhchem} 
\usepackage{cancel}
\usepackage{caption}
\usepackage{multirow}
\usepackage{subcaption}
\usepackage{algpseudocode}

\usepackage[normalem]{ulem}
\usepackage{color}
\usepackage{xcolor}

\title{Solving the Schr\"odinger Equation in the Configuration Space with Generative Machine Learning}

\author{Basile Herzog}
\affiliation[ULor]{Université de Lorraine and CNRS,
LPCT UMR 7019, F-54000 
Nancy, France}

\author{Bastien Casier}
\affiliation[ULor]{Université de Lorraine and CNRS,
LPCT UMR 7019, F-54000 
Nancy, France}

\author{S\'ebastien Leb\`egue}
\affiliation[ULor]{Université de Lorraine and CNRS,
LPCT UMR 7019, F-54000 
Nancy, France}

\author{Dario Rocca}
\affiliation[ULor]{Université de Lorraine and CNRS,
LPCT UMR 7019, F-54000 
Nancy, France}
\email{dario.rocca@univ-lorraine.fr}

\begin{document}

\begin{abstract}
The configuration interaction approach provides a conceptually simple and powerful approach to
solve the Schr\"odinger equation for realistic molecules and materials but is characterized by an unfavourable 
scaling, which strongly limits its practical applicability. Effectively selecting only the configurations that actually contribute to
the wavefunction is a fundamental step towards practical applications. 
We propose a machine learning approach that iteratively trains a generative model to preferentially generate the important configurations. 
By considering molecular applications it is 
shown that convergence to chemical accuracy can be achieved 
much more rapidly with respect to random sampling or the Monte Carlo configuration interaction  method. This work paves the way to a broader use of generative models
to solve the electronic structure problem.
\end{abstract}

	\maketitle

\section{Introduction}

The application of machine learning (ML) to quantum chemistry and computational materials
science has experienced an impressive growth in the past few years.
However, methods that apply ML to molecular dynamics~\cite{behler07,unke21}  
or molecular property predictions~\cite{von20, sanchez18} usually
imply the availability of a certain amount of data previously produced by approximating the solution of the Schr\"odinger equation. By considering the exponential numerical complexity 
involved in the exact solution of this equation
it would be highly desirable to take advantage of ML techniques also in this
context. For realistic molecules and materials this is a widely open field of research
and the full potential of ML has yet to be widely exploited in its full
potential.

In a seminal work Carleo and Troyer showed that neural networks  
can effectively represent the quantum states of spin models, thus
reducing the exponential complexity of the many-body problem \cite{carleo17}.
Specifically,
they showed that a restricted Boltzmann machine (RBM) used as an ansatz 
within variational Monte Carlo can achieve variational energies lower than those obtained with traditional approaches. From a theoretical point of view this success relies on universal approximation theorems \cite{hornik91,le08}, which imply that neural networks can approximate complex (but "reasonably" smooth) high-dimensional functions, including quantum states.
This method was demonstrated numerically considering the one- and two-dimensional Ising and Heisenberg models. More recently Carleo and 
coworkers have applied a similar approach to realistic Hamiltonians
of small molecules \cite{choo20}. This was achieved by mapping 
the electronic structure Hamiltonian into a spin-like Hamiltonian
by using quantum information encodings. The most accurate
results where obtained with the Jordan-Wigner mapping \cite{jordan28},
which leads to an approach equivalent to the configuration interaction (CI) \cite{szabo12}. Within the CI method
the fully interacting wavefunction is expressed as a linear combination of excited Slater determinants (the ``configurations''); while
the coefficients of this expansion are typically computed as solution of an eigenvalue problem, in Ref. \citenum{choo20} they were learnt by the RBM 
in an unsupervised way using a Monte Carlo sampling and 
the variational principle.
While this approach was achieving chemical accuracy for small basis sets,
the Monte Carlo sampling was repeatedly drawing the most dominant states (i.e. the Hartree-Fock determinant)
and this was at the origin of a significant slow down of convergence for basis sets beyond the minimal STO-3G.
It was recently shown that this issue can be alleviated by
using autoregressive neural networks, that allowed calculations
with up to 30 spin orbitals \cite{barrett22}. Our numerical results presented below
show that a simple RBM architecture used as a generative model can easily
double this number.
More in general it should be noticed that the representation of the wavefunction in the configuration space 
is strongly non-smooth (determinants with similar occupations can provide
significantly different contributions to the wavefunction) and this could be challenging
for ML regression approaches. To overcome this issue alternative methods 
use instead the high expressive power of deep neural networks
to represent electronic structure wavefunctions in real-space \cite{han19,hermann20,pfau20,kessler21}; as an advantage some of these techniques
can achieve the complete
basis set limit in a rather straightforward way but on the other side they require deep architectures involving the optimization of a large number of parameters and special care to keep into account the antisymmetry of the electronic wavefunction.

The antisymmetry is instead naturally included 
in the CI space which, however, grows unfavorably 
with the system size. 
This has lead to the development of a series of methods that select the excited determinants that
contribute the most to the wavefunction, either based on
perturbation theory \cite{huron1973iterative} or Monte Carlo sampling \cite{huron1973iterative}.
More recently machine learning techniques have been coupled to these methods \cite{coe18,pineda21,jeong21}. In this context Coe was the first to propose a Monte Carlo 
approach that explores the configuration space but, instead of explicitly
computing the coefficient of each configuration, this is predicted by a regression
neural network \cite{coe18}. While Carleo and 
coworkers apply a neural network approach to
``exactly'' fit the wavefunction in the configuration space
using the variational principle \cite{choo20},
the alternative approaches of Refs. \citenum{coe18,pineda21,jeong21} are more qualitative and
use supervised learning techniques.

\begin{figure}[!h]
    \centering
    \includegraphics[width=0.54\textwidth]{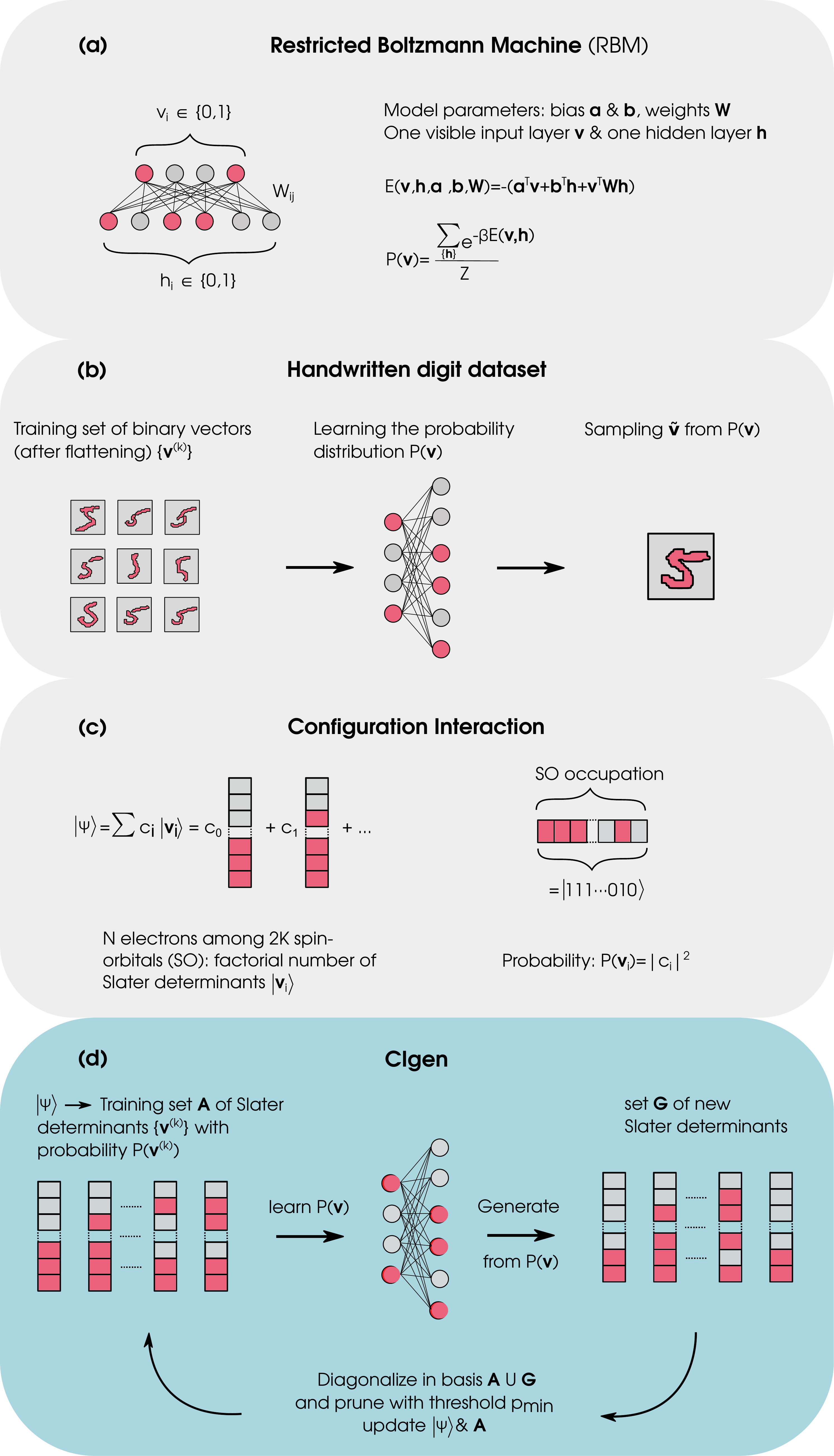}
    \caption{(a) Restricted Boltzmann machine (RBM) architecture consisting of one visible input layer and one hidden layer of binary values; for a given configuration ($\bm v,\bm h$) the parameters ($\bm a, \bm b, \bm W$) are used to define an energy function $E$ and an associated Boltzmann-like probability density $P$. (b) As an example, the RBM can be trained on a set of handwritten digits and afterward used to generate new realistic ones; to this purpose the digit's images are flattened to become unidimensional binary vectors
    $\mathbf{v}^{(k)}$ where 1 and 0 correspond to the digit and background pixels, respectively. (c) The configuration interaction (CI) approach expands the wavefunction of a molecule as a a linear combination of excited Slater determinants, which can be represented as a sort of unidimensional binary image. (d) The CIgen algorithm presented in this work iteratively trains an RBM on the distribution of determinants in the current approximation of the wavefunction and subsequently uses it to expand it by generating new important contributions.
    }
    \label{fig:TOC}
\end{figure}

In this work we propose an alternative approach named CIgen that uses generative machine learning to directly
generate the configurations that contribute the most to the electronic wavefunction. 
This is done with the general aim of avoiding the 
exploration of the overwhelmingly large configuration space
to search for the most important determinants. 
Notable examples of generative machine learning algorithms include the  restricted Boltzmann machines \cite{hinton06,melko19}, the variational 
autoencoders \cite{kingma13}, and the  
generative adversarial networks (GANs) \cite{goodfellow14}.
CIgen is based on RBMs, a type of generative neural 
network whose architecture is shown in Fig. \ref{fig:TOC}(a). As for other generative 
models the RBM can learn in an unsupervised way the statistical distribution
behind a series of input objects and then be used to generate new ones. 
A qualitative example is shown in Fig. \ref{fig:TOC}(b): An RBM can be trained
with a series of pictures, e.g. of handwritten digits, 
in practice represented as a flattened matrix of pixels and afterward
used to generate realistic new realistic images of the same digit.

For a given molecule or material the excited Slater determinants are associated
to an underlying probability distribution that could only be exactly evaluated
by solving the Schr\"odinger equation to obtain the wavefunction.
By representing the space
of excited determinants simply as binary vectors (a sort of
onedimensional binary image),
the RBM is here iteratively trained using data from the current approximation of the
wavefunction and subsequently used to generate new important determinants
to improve the wavefunction approximation (see Fig. \ref{fig:TOC}(c-d)).

Within the previous approach of Carleo and coworkers 
\cite{carleo17, choo20} 
the RBM was not properly used as a generative model but rather 
as a regression model for the wavefunction (to this purpose the RBM architecture was generalized to include
complex weights); here we use instead the RBM as a 
model to qualitatively represent
the probability distribution associated with the wavefunction, which is afterward used 
to generate the most likely configurations. In this respect the CIgen approach has some
similarities with the use of RBMs to learn the statistical distribution of data from experimental measurements of quantum states
and to subsequently generate new configurations for quantum averages\cite{melko19}.
Differently from
selective CI or Monte Carlo CI our approach does not ``explore'' the huge
determinant space to find significant contributions but rather 
directly generate them with the current approximation of the probability
distribution associated with the RBM.
Comparison with previous approaches will be further discussed below.

\section*{Results and discussion}

\subsection*{Configuration interaction approach}

Considering the non-relativistic N electron problem in the Born-Oppenheimer approximation, the configuration interaction method solves the Schrödinger equation for a fixed basis set by expanding the wavefunction in the following way \cite{szabo12}:
\begin{equation}
    |\Psi \rangle = c_0  |\Phi_0 \rangle
    +\sum_{r a} c_a^r |\Phi_a^r \rangle
    +\sum_{\substack{a<b \\ r<s}} c_{a b}^{r s} |\Phi_{a b}^{r s} \rangle
    +\sum_{\substack{a<b<c \\ r<s<t}} c_{a b c}^{r s t} |\Phi_{a b c}^{r s t} \rangle
    +\cdots 
\label{fullci}
\end{equation}
where $|\Phi_0 \rangle$ is the Hatree-Fock ground state, $|\Phi_a^r \rangle$ 
is a singly excited Slater determinant from occupied spin orbital $a$ to unoccupied spin orbital $r$, and all the other terms
correspond to multiple excitations; while $| \Psi \rangle $ could denote any excited
state of a given system here the discussion will be focused on the ground-state. 
When all the possible excited determinants are included in Eq. \ref{fullci}
the approach is called full configuration interaction (FCI) and the solution
of the Schr\"odinger equation becomes exact for a given basis set.
Within the ``brute-force'' FCI approach  
the coefficients of the expansion
$c_0, c_a^r, \cdots$ are determined by computing the
expectation value of the Hamiltonian $\langle \Psi | \hat{H} | \Psi \rangle$ and
applying the variational principle. 
Since the total number of determinants grows as $\binom{2K}{N}$, where $2K$ is the
number of spin-orbitals, the FCI approach becomes
quickly unpractical for most applications. It is however well known that
often the wavefunction 
can be accurately represented by a limited number of excited determinants and 
here a ML model is proposed to directly generate the important contributions. \\

\subsection*{Generative model}

In this work, it is shown that restricted Boltzmann machine (RBM) can be used to efficiently 
generate the excited determinants that contribute the most to the wavefunction in Eq.
 \ref{fullci}. An RBM, pictorically represented in Fig. \ref{fig:TOC}, is a generative neural network consisting of one input layer of $D$ visible binary units $\{v_i\} \in \{0,1\}^D$, one layer of $P$ binary hidden units $\{ h_j \} \in \{0,1\}^P$, and $D \times P$ weights $\{ W_{ij} \}$ between both layers. Two bias vectors $\{ a_i \}$ and $\{ b_j \}$ are added to the visible and hidden layer, respectively. \\
By introducing an energy function of $\Lambda = \{ \bm a, \bm b, \bm W \}$ for a given configuration $\{ \bm v, \bm h \}$ as
\begin{equation}
    E(\bm{v}, \bm{h}, \Lambda) = - (\bm{a}^T \bm{v} + \bm{b}^T \bm{h} + \bm{v}^T \bm{W} \bm{h}),
    \label{Erbm}
\end{equation}
and an inverse temperature $\beta =1/T$, one can define the probability distribution associated with the RBM over the input configurations $\bm v$ to be
\begin{equation}
    P(\bm v) = \dfrac{ \sum_{ \{ \bm h \}} \mathrm{e}^{-\beta E(\bm v, \bm h)}}{\sum_{ \{ \bm h, \bm v \}} \mathrm{e}^{-\beta E(\bm v, \bm h)}}.
\label{proba}
\end{equation}
After training on a set of vectors  $\bm v$
distributed according to $P(\bm v)$, the RBM can be used to
generate new vectors $\bm v$ according to this
probability distribution.
To this purpose the Gibbs sampling can be used \cite{mackay2003information}: Starting from an initial
random trial vector as input, one can obtain a
hidden-layer vector $\bm h$ using the conditional probability $p(h_i | v_i)$ and then obtain a new input-layer vector $\bm v$ using the conditional probability $p(v_i | h_i)$; the repetition of this operation 
a certain number of times forms a Markov chain that
generates a vector $\bm v$ according to
the probability $P(\bm v)$.

For the purpose of this work the input vectors $\bm v$
represent the Slater determinants 
and their associated 
probability $P(\bm v)$ should ideally be proportional
to their contribution to the wavefunction (the square
of the $c$ coefficients in Eq. \ref{fullci}). 
The determinants are represented as binary vectors, where 1 denotes occupied and 0 unoccupied states, with size equal to the number of spin-orbitals
(see Fig. \ref{fig:TOC}). 
The architecture of the RBM used in this work is rather
simple, with 2K (the number of spin orbitals) neurons
in the visible and in the hidden layers.
Given a certain fixed subspace of determinants,
the Hamiltonian is diagonalized to obtain    
the wavefunction coefficients; the model is then trained with random configurations generated according to the corresponding probability distribution by 
updating the RBM parameters with the contrastive divergence algorithm \cite{hinton2012practical}. 
The training and generative procedure of an RBM are discussed in detail in Supplementary Material.
The generation of new determinants and the retraining of the model are performed iteratively.
Specifically, the CIgen algorithm performs the following steps:
\begin{enumerate}
    \item Start: The configuration interaction singles and doubles (CISD) is used to generate an
    initial guess wavefunction.
    \item Pruning: The determinants whose squared wavefunction coefficients (Eq. \ref{fullci}) are below a certain threshold $p_{min}$ are pruned.
    \item Training: The RBM model is trained only on the non pruned determinants but the Hartree-Fock determinant, which is typically associated to a very high probability, is not included in the training set (otherwise the ML would mainly generate the Hartree-Fock state in the following step). 
    \item Generation: The trained RBM is used to generate a set of new determinants, whose number is proportional to the number of determinants already included in the current approximation of the wavefunction. During this procedure determinants are automatically discarded if they are already included in the wavefunction, if they do not couple through single or double substitutions with the current determinant set, and if they do not have correct spin and point group symmetries.
    \item Diagonalization: The Hamiltonian is diagonalized in the subspace that includes the newly generated determinants and new coefficients are obtained for Eq. \ref{fullci}.  
    \item Iterate: The procedure is repeated from step 2 until convergence of the energy is achieved.
\end{enumerate}

Differently from previous  
approaches based on RBMs for quantum states \cite{carleo17, choo20}, CIgen  does not require an RBM architecture generalized
to include complex weights and, while the model still learns in an unsupervised way, the procedure is not based on the
variational principle. 
While the training of a ML model based on the variational principle is 
more elegant and physically motivated \cite{carleo17, choo20}, the CIgen approach is less prone to 
overfitting as it does not aim to exactly fit the wavefunction.

\if
CIgen presents several differences with respect to previous traditional 
approaches based on RBMs for quantum states \cite{carleo17, choo20}: (1) While previous
approaches use RBMs with complex weights to represent the wavefunction, here the
RBM learns the probability distribution associated with the wavefunction and
uses real weights; (2) Our RBM model learns in an unsupervised way but does not use 
the variational principle of quantum mechanics; (3) The RBM is not used as an ansatz
for the variational quantum Monte Carlo but at every iterative step is used
to generate a set of new determinants and retrained with the new data.
While the training of the ML model based on the variational principle is 
more elegant and physically motivated cite{carleo17, choo20}, this procedure is less effective in the
Slater determinant space, where the uneven probability landscape makes regression
more difficult. Our approach avoids this issue by using the RBM to learn the
wavefunction probability distribution only qualitatively with the purpose of
generating new determinants.
\fi


\subsection*{Direct generation of single and double excitations}

The CIgen algorithm is based on a rather straightforward way of using a generative neural network model to enlarge the configuration subspace of interest. 
However, the generative step (step 4 of the CIgen algorithm) often
samples determinants that clearly cannot contribute to the wavefunction
and this makes the overall procedure less efficient. 
For example, it is well known that a fixed 
set of determinants can only be coupled through the Hamiltonian to 
new determinants that differ by single and double substitutions.
This observation, which has some subtle implications, is exploited by the Monte Carlo Configuration Interaction (MCCI) algorithm proposed by Greer \emph{et al.} \cite{greer1998monte}.
In the MCCI algorithm, new determinants are sampled by randomly proposing 
single and double excitations of the determinants already included in the wavefunction
set. At first sight this could seem a rather inefficient random procedure but in practice 
the important (highly contributing) determinants have often multiple connections
to other important determinants through single and double substitutions and 
are much more likely to be sampled. To take advantage of this feature of the
configuration space we developed a variant of our approach named CIgen-SD.
Within this procedure a determinant $\Phi_1$ is first randomly selected among those included in the current 
approximation of the wavefunction. Through the RBM a new determinant $\Phi_2$ is then sampled and used
exclusively
to construct a transition probability matrix.
Finally, a new determinant is proposed by exciting $\Phi_1$ according to this probability matrix
and the whole procedure is repeated several times.
The CIgen-SD procedure is explained in details in Supplementary Material.
The CIgen, CIgen-SD, and MCCI will be compared in the Numerical Results subsection.

\subsection{Numerical Results}

We now discuss the efficiency and accuracy of the CIgen approach
by considering applications to molecular systems. 
To this purpose total energy calculations for the C$_2$, N$_2$, and H$_2$O molecules are performed for the 6-31G and cc-pVDZ
basis sets considering the active spaces
indicated in the third column of Table \ref{tab:RBMen}. For the largest system
the number of spin orbitals is about double than what
previously achieved with neural network quantum states \cite{barrett22}.

As a preliminary step it is important to discuss the effect of the temperature
on the generative power of the RBM (see Eq. \ref{proba}). Fig. S1 shows the convergence of
the total energy of H$_2$O in the cc-pVDZ for three values of the temperature.
The convergence of the CIgen approach is optimal at T=1 (chemical accuracy is achieved within  11 iterations) but sizeably slows down 
when T is increased to 5. In the limit of a very large temperature (T=500) the method
performs poorly. Indeed, as explained in SM, the 
$T \to \infty$  limit leads 
to a completely random generation of determinants.
This shows that for reasonable choices of the temperature the CIgen method actually learns the 
probability associated with the wavefunction and significantly speeds up
the generation of the excited determinants with respect to random sampling. \\

We now consider the convergence of the total energy in the cc-pVDZ basis for the three molecules C$_2$, N$_2$, and H$_2$O in the cc-pVDZ basis and compare the perfomance of the CIgen and CIgen-SD algorithms with the MCCI algorithm. 
In all three cases the two CIgen variants considered here outperform the MCCI method. This is particularly true for the CIgen-SD approach that
achieves chemical accuracy with a significantly smaller number of iterations.

\begin{figure}[h!]
    \centering
    \includegraphics[width=0.7 \textwidth]{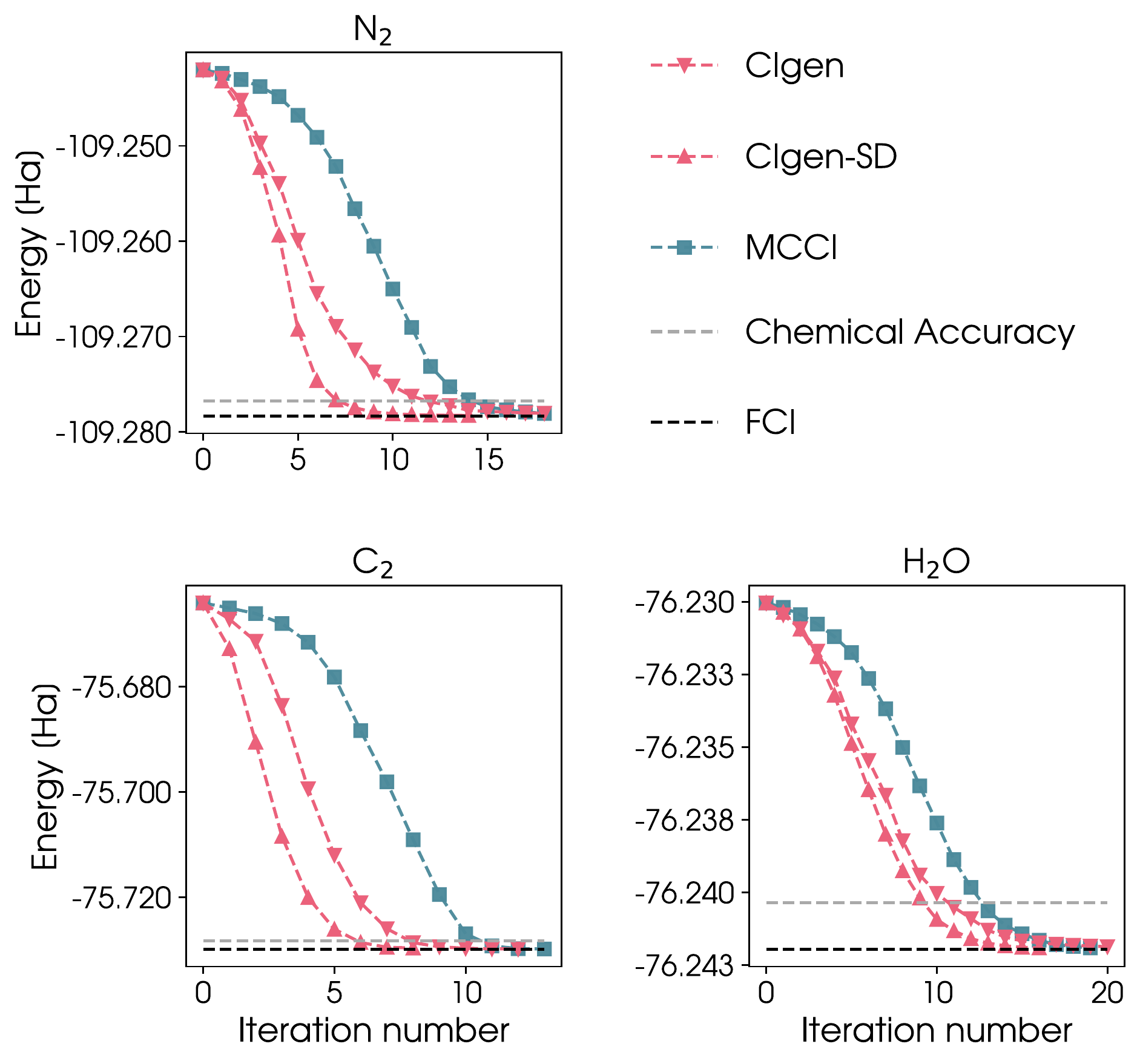}
    \caption{Total energy convergence for N$_2$, C$_2$, and H$_2$O in the cc-pVDZ basis set as a function of the number of iterations in the CIgen, CIgen-SD and MCCI algorithms. The full configuration interaction (FCI) reference values and the chemical accuracy threshold are represented by horizontal lines.}
    \label{fig:allc}
\end{figure}

The detailed results for the total energies of the molecules considered here are 
presented in Table \ref{tab:RBMen}, where CIgen-SD values are compared to
coupled-cluster with singles, doubles and perturbative triples CCSD(T) and to FCI
reference values.
By considering a limited number of iterations (from
a minimum of 9 for C$_2$ in 6-31G to a maximum of 17 for water in cc-pVDZ),
CIgen-SD converges to FCI values within chemical accuracy.
This is achieved by generating an excited determinant
subspace that is significantly smaller than the full space.
For example, for the molecule N$_2$ in the cc-pVDZ basis set  only 3.5 million determinants are
generated and included in the wavefunction out of the about 540 millions
that would be allowed by spin and space symmetries.
This significant reduction of two orders of magnitude in the number of determinants
is not achieved by a search in the configuration space but rather by 
an iterative generation (and model training) of the determinants that contribute the most to the total energy.



\begin{table}[htpb!]
\centering

\begin{tabular}{||c|c|c|c|c|c|c||}
\hline
System         &   Basis & (N,K)   & E$_{CIgen-SD}$ & E$_{CCSD(T)}$ & E$_{FCI}$ & N$_{conv}$   \\
\hline
C$_2$ & 6-31G & (12,18) & -75.64416 & -75.64415 & -75.64418 & 12\\
\hline
N$_2$ & 6-31G & (14,18) & -109.10824  & -109.10635 & -109.10842 &  12\\
\hline 
H$_2$O & 6-31G & (10,13) & -76.12220 & -76.12182 & -76.12237 & 10\\
\hline
C$_2$ & cc-pVDZ & (8,26) & -75.72982  & -75.72781 & -75.72984 & 9\\
\hline
N$_2$ & cc-pVDZ & (12,27) & -109.27827 & -109.27829 &-109.27834 & 13\\
\hline 
H$_2$O & cc-pVDZ & (8,23) & -76.24192 & -76.24131 &-76.24195 & 17\\
\hline
\end{tabular}
\caption{\label{tab:RBMen} Total energies (Ha) for the different systems/basis sets considered here as obtained with the CIgen-SD method; reference FCI and CCSD(T) energies are provided for comparison purposes. The (N,K) column indicate the size of the active space,
where N is the number of correlated electrons and K is the number of active molecular orbitals, and the N$_{conv}$ column indicates the number of iterations to achieve convergence (i.e
when the energy in two successive iterations differs by less than $10^{-5}$ Hartree).}
\end{table}





\begin{figure}[h!]
    \centering
    \includegraphics[width=0.7 \textwidth]{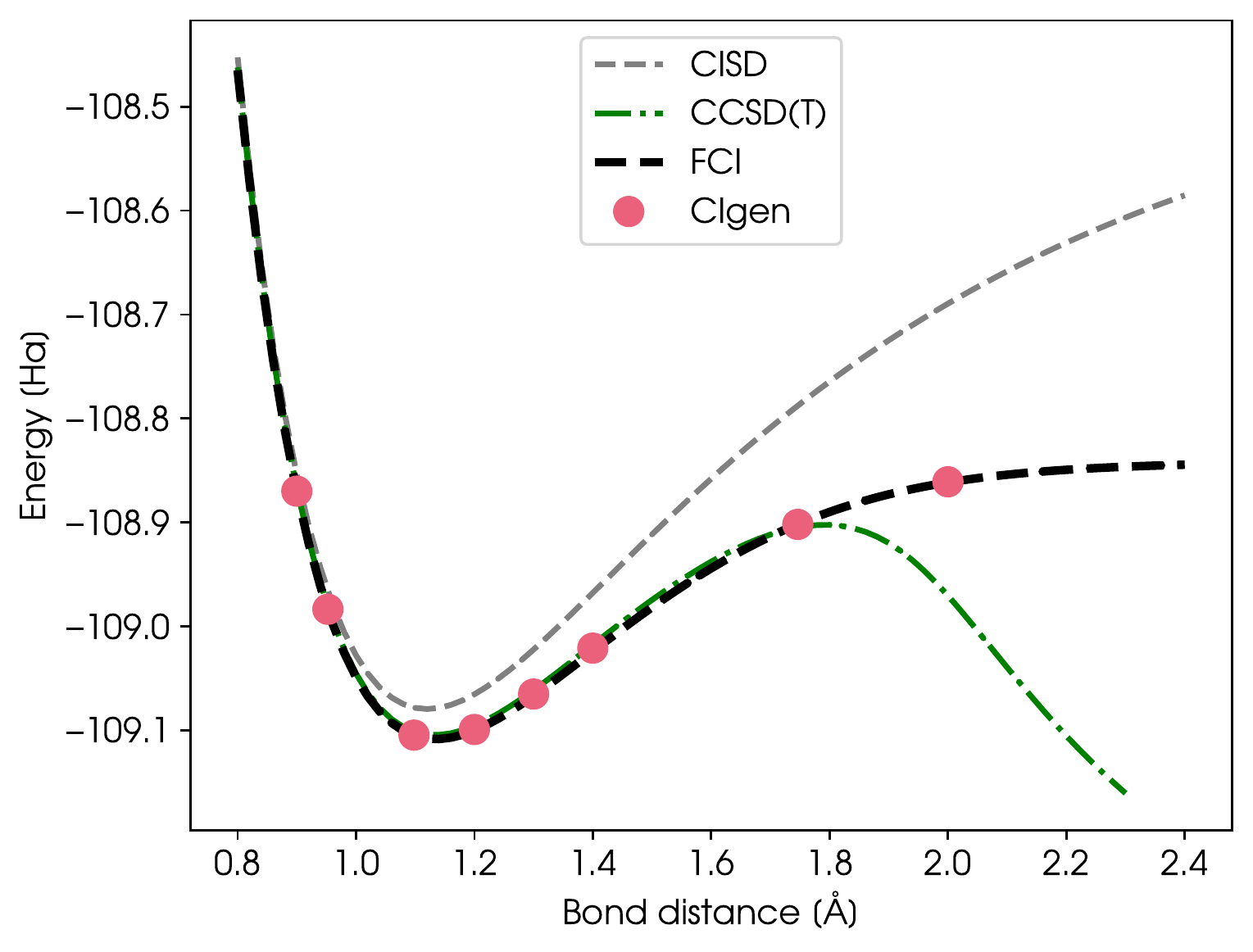}
    \caption{CIgen dissociation curve of N$_2$ in the 6-31G basis set compared to the CCSD(T), FCI, and CISD curves.}
    \label{fig:bindingN2}
\end{figure}

As a final result in Fig. \ref{fig:bindingN2} we present a full
binding curve for the N$_2$ molecule. This represent a much more challenging example
since for large interatomic distances the binding is 
characterized by strong static correlation whose description could become problematic for
several traditional quantum chemical approaches.
This is the case of CCSD(T) that starting at around 1.8 \AA \ produces a
binding curve with an unphysical behavior.
The CIgen approach well describes the N$_2$ binding curve at every distance and
produces a curve in excellent agreement with FCI. 
The comparison with the CISD curve, which provides the starting training data for 
the generative model, shows clearly that CIgen recovers an increasing amount of
electronic correlation energy as a function of the interatomic distance.

In summary we have shown how a generative model can be used to solve the Schr\"odinger equation
by sampling the excited Slater determinants that contribute the most to the wavefunction.
Numerical applications show that this approach is already competitive with previous approaches based on Monte Carlo sampling of the excited determinants \cite{greer1998monte}
or on machine learning ansatzes to represent the wavefunction in the configuration space
\cite{choo20,barrett22}.
An improvement that should be addressed in future work involves the development
of a generative model that intrinsically takes into account the symmetry of the determinants and,
more in general, the properties of the determinants contributing to the wavefunction. 
It has already been shown that a sizeable improvement in the convergence speed can be achieved  by directly generating the determinants that couple with double and single
substitutions to the determinants already included in the wavefunction.
In the current implementation, however, spin and spatial symmetries of the generated determinants are
verified and enforced only \emph{a posteriori}. The development of a generative machine learning
model whose architecture keeps into account these symmetries would certainly increase the 
numerical efficiency of CIgen to possibly perform calculations with significantly larger numbers of
spin orbitals.

\section{Methods}

\subsection{Computational details}

The equilibrium geometries of the molecules considered in this work are optimized at the
CCSD(T) level of theory in the corresponding basis set (6-31G or cc-pVDZ) and
can be obtained from the Computational Chemistry Comparison and Benchmark DataBase \cite{nist}.
The reference results for these molecules at the CCSD(T) and FCI level of theory
(Table \ref{tab:RBMen} and Figs. \ref{fig:allc}-\ref{fig:bindingN2}) have been
obtained from the Molpro code \cite{MOLPRO_brief, KH89}.

The electronic structure part of our implementation (specifically the 
diagonalization of the Hamiltonian in the space of the generated determinants)
is based on the Quantum Package code \cite{Garniron_2019}.
The selection of the determinants to be included in the wavefunction in Eq. \ref{fullci}
is based on an acceptance threshold of
$10^{-12}$ on the squared coefficients.

\begin{acknowledgement}
The authors would like to warmly thank Mauricio Chagas da Silva for valuable discussions, as well as Anthony Scemama for his help with Quantum Package.
This work was supported through the COMETE project (COnception in silico
de Mat\'eriaux pour l'EnvironnemenT et l'Energie) co-funded by the European 
Union under the program ``FEDER-FSE Lorraine et Massif des Vosges 2014-2020''. B.H., S.L., and D.R. acknowledge the financial support of the Agence Nationale de la Recherche under the Lorraine Artificial Intelligence (LOR-AI) project (grant number ANR-20-THIA-0010-01). 
\end{acknowledgement}

    
    

	\bibliography{refs}

\end{document}